\documentclass[
  journal=pasa,
  manuscript=research-paper, 
  year=2025,
  volume=YY,
]{cup-journal}

\usepackage{amsmath}
\usepackage{amssymb,microtype,siunitx,booktabs}

\usepackage{soul}

\newcommand{\qlim}{q_\mathrm{lim}}
\newcommand{\fbin}{f_\mathrm{bin}}
\newcommand{\msol}{\mathrm{M_\odot}}

\title{Rubin Data Preview 1: Extending the View of Unresolved Binary Stars in 47\,Tucanae}

\author{G. Cordoni}
\affiliation{Research School of Astronomy and Astrophysics, The Australian National University, Canberra, ACT 2611, Australia}
\email[G. Cordoni]{giacomo.cordoni@anu.edu.au}

\author{L. Casagrande}
\affiliation{Research School of Astronomy and Astrophysics, The Australian National University, Canberra, ACT 2611, Australia}

\author{H. Jerjen}
\affiliation{Research School of Astronomy and Astrophysics, The Australian National University, Canberra, ACT 2611, Australia}

\received {1 August 20  25}
\revised  {dd Mmm YYYY}
\accepted {dd Mmm YYYY}
\published{dd Mmm 2025}


\begin{document}

\begin{abstract}
Until now, the study of unresolved main sequence binary stars in globular clusters 
has been possible almost exclusively in their central regions with deep Hubble Space Telescope (HST) observations. We present the first detection of unresolved main-sequence binary stars in the outer field of 47\,Tucanae using Rubin Observatory’s Data Preview 1 (DP1). Our analysis exploits deep $i$ vs. $g-i$ colour–magnitude diagrams beyond the cluster's half-light radius, reaching almost to the tidal radius. The high-quality photometry allowed to identify unresolved binaries with mass ratios $q$ larger than 0.7. The derived binary fraction of $\fbin (q>0.7)=0.016\pm0.005$ stands in contrast to the significantly lower values in the cluster innermost regions, as measured from HST photometry.

This result provides new empirical input for testing physical processes that drive the formation and evolution of binary stars in globular clusters. It also demonstrates Rubin’s unique wide-field and high-precision photometric capabilities to address a broader range of outstanding questions in star cluster research. Future full data releases will enable to significantly expand the study of dense stellar systems across the Milky Way. 
\end{abstract}

\section{INTRODUCTION }
\label{sec:int}

The Legacy Survey of Space and Time \citep[LSST][]{ivezic2019, usher2023} will provide deep and wide-field imaging of the southern sky using the $ugrizy$ photometric bands. Before starting the full survey, the Rubin Observatory has released preliminary, science-grade data through its Early Science Program, beginning with Data Preview 1 \citep[DP1, ][]{rubinDP1dataset2025}. DP1 includes calibrated single-epoch images, coadded images, difference images, and corresponding source catalogues. One of the DP1 fields is centred on the globular cluster (GC) 47 Tucanae (47\,Tuc), offering the first opportunity to test Rubin's performance in dense stellar environments.

In this work, we utilise Rubin DP1 data \citep{rubinDP1dataset2025} to investigate the unresolved main-sequence binary fraction in the outskirts of 47\,Tuc, cluster regions that were previously difficult to study due to observational limitations. Until recently, such studies were mainly carried out using Hubble Space Telescope (HST) or James Webb Space Telescope (JWST) data for the dense GC cores \citep[e.g.][]{bellazzini2002a, richer2004a, sollima2007, dalessandro2011, milone2012a, milone2016a, milone2020, bortolan2025} and Magellanic Clouds clusters \citep{mohandasan2024, muratore2024} or using Gaia data for Galactic open clusters, which are brighter and less dense environments \citep[e.g.][]{cordoni2023, donada2023}.

Binary stars play a crucial role in stellar and cluster evolution and dynamics, for example by affecting energy exchange, driving mass segregation, and influencing core collapse and evaporation timescales \citep{hut1992, gill2008}. Observationally, unresolved binaries affect luminosity and mass functions, leading to systematic errors in cluster total‐mass and mass‐function estimates \citep{bianchini2016}.

Traditional methods for identifying binaries, by means of radial velocity or photometric variability, typically favour brighter stars or binaries with short orbital periods. In this work, we adopt a photometric method that exploits the different position of unresolved main-sequence binaries in colour-magnitude diagrams (CMDs), with respect to single main-sequence (MS) stars. In CMDs, binaries typically appear brighter and redder than single stars shifting them to a locus parallel to the MS\citep[][]{bellazzini2002a, richer2004a, sollima2007, milone2012a}. This photometric approach requires observations in at least two photometric filters and allows the study of large numbers of stars simultaneously.

\section{DATA}
\label{sec:data}
We use photometric data from the Rubin Observatory’s DP1, collected with the Rubin Commissioning Camera \citep[LSSTComCam][]{lsstComCam2024}. DP1 comprises more than 1,700 science-grade exposures gathered in late 2024 across multiple bands ($ugrizy$), processed using the LSST Science Pipelines \citep{developers2025}. These preliminary observations aim to assess and refine Rubin Observatory's systems and provide early data access for scientific evaluation \citep{guy2025}.

Our analysis focused on the globular cluster 47\,Tuc at a distance of $\sim$4.5\,kpc, covering a region extending outside the cluster's half-light radius \citep[$R_\mathrm{h}=3.17'$][revision of 2010]{harris1996}, between approximately 18 and 40\,arcmin from its centre. Data have been downloaded from the Rubin Science Platform \citep[][]{juric2019}\footnote{\url{https://lse-319.lsst.io/}}, using a similar query to \citet[][]{choi2025}, selecting data from the $gri$ bands. Unfortunately, $u$ band visits failed to meet the Rubin DP1 internal quality criteria, and were not publicly released. We refer to \citet{choi2025} for a detailed description of the number of visits, limiting magnitude and seeing distribution in the different bands.

\begin{figure*}
    \centering
    \includegraphics[width=0.8\linewidth]{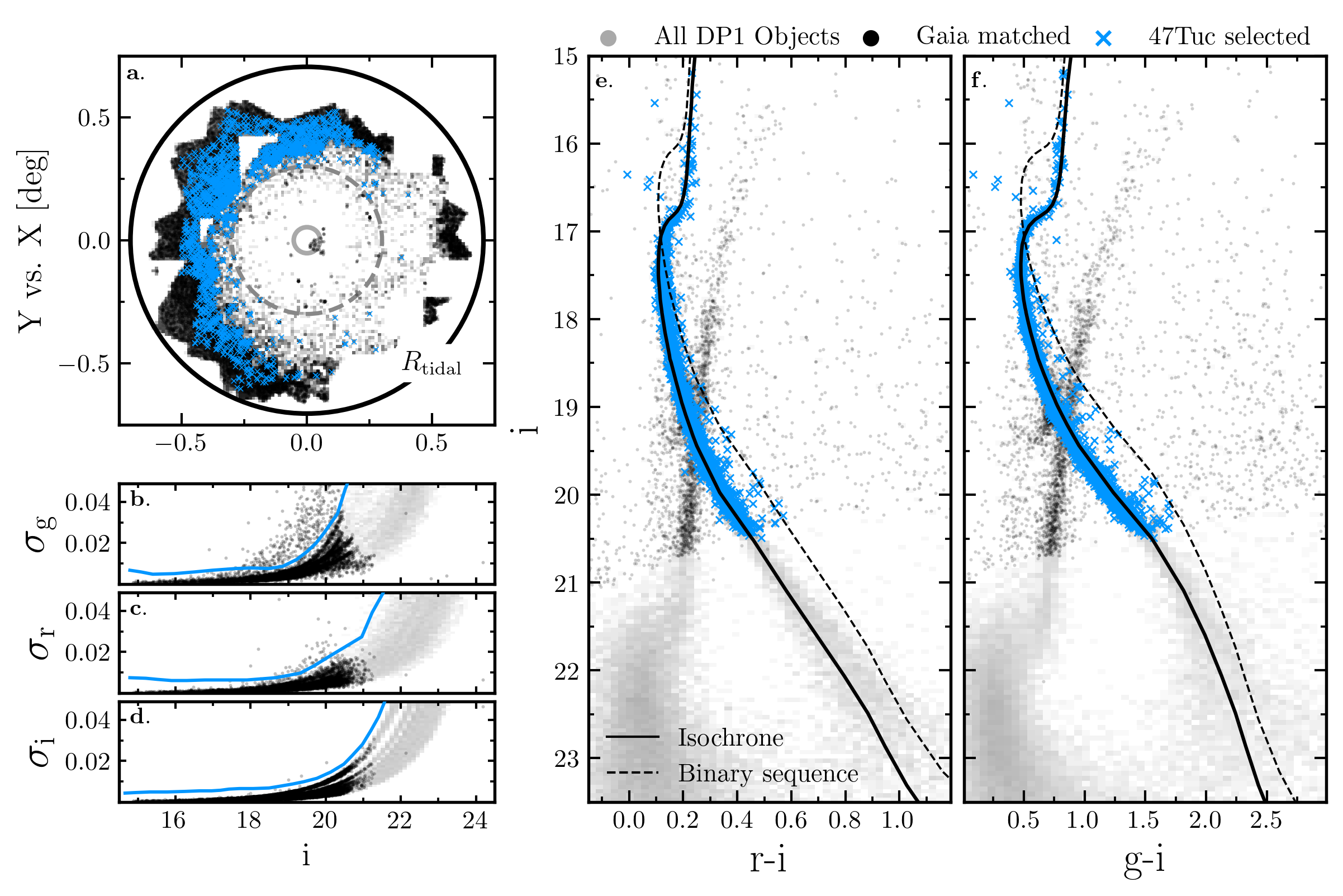}
    \caption{Data selection procedure. \textit{Panel a.} Distribution of stars in the whole sample of Rubin DP1 coadd object catalogue (grey), stars with a Gaia matched source (black) and 47\,Tuc selected cluster members (azure). X and Y have been determined projecting ra and dec and are displayed in degrees, while the solid grey and black circles represent the half-light and tidal radii, respectively. The dashed grey circle indicate the 18\,arcmin radius utilised in the selection of the final sample of stars. \textit{Panels b-d.} Uncertainties on $gri$ photometry as a function of $i$ magnitude. The azure lines indicate the threshold adopted to select stars with low uncertainties. \textit{Panel e-f.} $i$ vs $r-i$ and $g-i$ CMDs. Colour-coding as for previous panels. The best‐fit PARSEC isochrone \citep{marigo2017a} as in \citet[][]{choi2025} is plotted as a solid black line, and the equal‐mass MS binary sequence as a dashed black line.}
    \label{fig:fig1}
\end{figure*}

As the objective of this work is to determine the unresolved binary fraction, we require a clean sample of 47\,Tuc stars. To this goal, we apply different quality cuts to the complete coadd DP1 photometric catalogue. Our selection criteria prioritised quality and purity, thus we relied exclusively on coadd photometry from stacked visits rather than forced photometry on individual visits. 
While the latter would have increased the sample size \citep{wainer2025}, it would also have introduced more contamination and lower-quality measurements, undesirable for our analysis.
To select well measured cluster members we adopted the following criteria:
\begin{itemize}
    \item We excluded sources with a \texttt{refExtendedness=1} flag, removing objects flagged as extended or non-stellar from the Rubin pipeline \citep[][]{bosch2019}.
    \item Stars with \texttt{iPSF=True} in any of the $gri$ bands have been removed to eliminate detections impacted by bad pixels.
    \item Excluded sources with \texttt{blendedness} in $gri$ bands larger than 0.05 to avoid sources possibly affected by photometric blends (see Section~\ref{subsec:blends} for more details). 
    \item Cluster membership was then determined through crossmatching with Gaia Data Release 3 \citep[DR3, ][]{gaiaDR3}, and adopting the selection criteria of \citet{vasiliev2021} and using stars with membership probabilities above 0.8.
    \item Adopting metallicity, distance and reddening as in \citet[][]{choi2025}, i.e. $[M/H]=-0.5;d=4.6~\mathrm{kpc};\,E(B-V)=0.025$, we excluded stars with colours differences larger than $\pm 0.4$ from the best-fit PARSEC isochrone \citep[][black lines in Figure~\ref{fig:fig1}e, f]{marigo2017a}.
\end{itemize}

The selection procedure is illustrated in Figure\,\ref{fig:fig1} where we show the whole Rubin DP1 coadd object catalogue in grey, Gaia matched sources in black, and selected sources as azure crosses. In addition to the discussed selection, we also excluded stars with large $gri$ photometric uncertainties, as shown in Figure\,\ref{fig:fig1}b-d where the azure lines indicate the adopted thresholds. Limits on photometric uncertainty were defined by computing the median uncertainty and its standard deviation within 10 equally populated magnitude bins, and adopting as threshold the median trend shifted upward by $3\sigma$. Finally, we removed stars within 18 arcmin from the cluster centre where crowding is stronger.

The complete coadd Rubin photometric catalogue includes 47\,Tuc stars, Small Magellanic Cloud (SMC) and field stars. Indeed, the CMD of the SMC is clearly visible extending from the bottom-left corner (MS stars) to the upper right (RGB stars) of Figure~\ref{fig:fig1}e and f. However, our selection criteria allow us to define a high-fidelity sample of 47\,Tuc stars, extending from the base of the RGB down to MS stars of approximately $0.5\,\msol$. The resulting CMD reveals a narrow, well-defined MS, with MS–MS binary stars clearly separated from single MS stars. Figures~\ref{fig:fig1}e and f further show that SMC stars are correctly removed by our selection criteria.

While 47\,Tuc's\,CMD is clearly distinguishable down to very faint magnitudes ($i\sim 22$), the main limiting factor is the depth of Gaia photometry, which only allows to select stars brighter than $i\sim 20.5$. Gaia incompleteness depends primarily on apparent magnitudes, which affect equally single and binary stars. Future Rubin data releases will also allow to constrain spatial incompleteness across the cluster.

The zoomed-in CMD in Figure~\ref{fig:fig2}a clearly demonstrate the good photometric quality and absence of significant differential reddening \citep[consistent with][]{legnardi2023, pancino2024}, which would otherwise spread the MS along the reddening direction, depicted as a grey arrow.

\section{RESULTS}
\label{sec:discussion}

To estimate the fraction of unresolved main-sequence binary systems, we utilise the method introduced in \cite{richer2004a, bellazzini2002a, sollima2007, milone2012a} and also  successfully applied to different classes of stellar systems  \cite[see e.g.][]{donada2023, cordoni2023, mohandasan2024, muratore2024}. In a nutshell, unresolved MS-MS binaries appear redder and brighter than single MS stars in the CMD. 
More specifically, their location in the CMD depends on the luminosity or mass of the primary star, and the mass-ratio $(q = M_\mathrm{sec} / M_\mathrm{prim}$. Equal-mass binaries $(q = 1)$ lie approximately 0.75 mag above the MS (dashed lines in Figure~\ref{fig:fig1}e, f), whereas binaries with smaller $q$-values approach the MS fiducial line. 

\begin{figure*}
    \centering
    \includegraphics[width=0.8\linewidth]{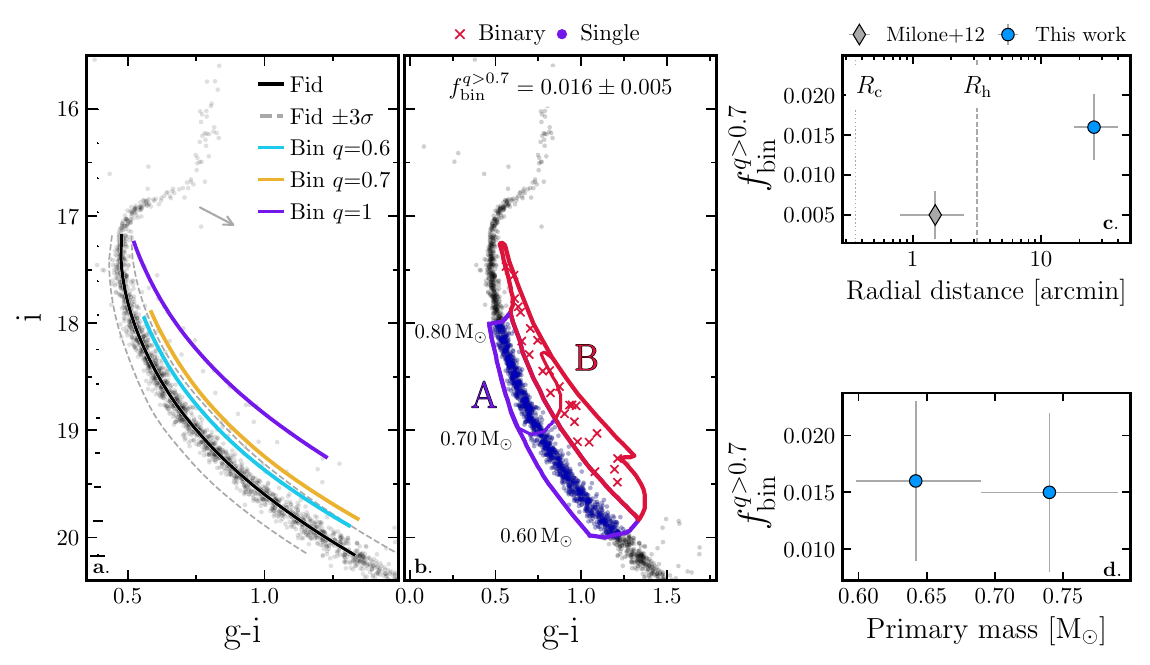}
    \caption{Binary fraction determination. \textit{Panel a.} $i$ vs. $g-i$ CMD of selected cluster members. The black line represent the fiducial line of the MS, while the grey lines are shifted by $\pm 3\sigma$, where $\sigma$ is the spread in colour at each $i$ magnitude. The fiducial line of MS-MS binary with mass-ratio of $q=0.6;0.7;1$ are shown in light-blue, orange and purple, respectively. Photometric uncertainties are displayed on the left side with black errorbars. \textit{Panel b. } Same CMD as panel a, with the regions of single and binary stars shown in purple (region A) and red (region B), respectively. Identified single and binary stars are marked with purple circles and red crosses. \textit{Panel c.} Recovered binary fraction in this work, compared with the value determined in \cite{milone2012a} from HST photometry of the cluster centre. Dotted and dashed vertical line indicate the core and half-light radii, respectively. \textit{Panel d.} Binary fraction for different high and low mass primary mass stars, determined in the two regions depicted in the CMD on panel b.}
    \label{fig:fig2}
\end{figure*}

In our analysis, we set a lower limit mass-ratio, $q_\mathrm{lim} = 0.7$ (red line in Figure~\ref{fig:fig2}a), to unambiguously distinguish binaries from single MS stars. This threshold was determined by computing the fiducial line of MS stars, and applying a colour shift of $3\sigma$, where $\sigma$ is the colour spread associated to the fiducial line at any given magnitude. The smoothed fiducial line and the $3\sigma$ shifted borders are displayed in Figure~\ref{fig:fig2}a as black and grey lines, respectively. The fiducial and the spread have been determined by computing the colour median and dispersion in magnitude bins of width 0.5 mag and equally spaced by 0.25 mag. As can be seen, binary systems with mass-ratios lower then 0.7, e.g. $q=0.6$ shown with the orange line, would overlap with MS stars. Therefore, $q_\mathrm{lim} = 0.7$ is the lowest mass-ratio for which we can reliably disentangle MS and MS-MS binary stars. We also include nominal photometric uncertainties as black errorbars on the left hand side of Figure~\ref{fig:fig2}a.

To determine the fraction of binaries we first identified a magnitude range where MS stars are clearly separated from binaries, i.e.~$i=18, 20$ as bright and faint magnitude limits. We then used PARSEC isochrones \citep[][]{marigo2017a} to compute the fiducial line of binaries with a mass-ratio equal to the selected $\qlim=0.7$. This marks the border between the region of MS single stars, having colours between the fiducial line shifted bluewards by $3\sigma$ and the fiducial line of binaries with $q=0.7$. Binary stars are instead selected as stars with colours between the fiducial of $q=0.7$ and the fiducial line of equal mass binaries shifted by $3\sigma$. The procedure is illustrated in Figure~\ref{fig:fig2}a,b. Specifically, the region including single stars is denoted as region A, while region B indicates the locus of binary stars. Single and binary stars are highlighted with blue circles and red crosses, respectively. Additionally, we subdivided regions A and B into two mass intervals ($0.6<M/M_\odot<0.7$, $0.7<M/M_\odot<0.8$ that contain approximately the same number of stars, allowing us to explore the dependence of binary fraction on primary mass\footnote{Region B have been split by connecting the colours and magnitude of binary systems with primary mass $M_\mathrm{prim}=0.7$ and $q$ varying from 0.7 to 1, as in \citet[][]{dalessandro2011, milone2012a, cordoni2023}.}. 
Finally, the fraction of binary stars is determined as the ratio between the number of binary candidates (stars in region B) and the total number of stars (sum of the single and binary candidates in regions A and B). 

\subsection{Estimating Field Star Contamination}\label{subsec:field}
One possible source of uncertainty is residual field stars contamination. ndeed, even after adopting the cluster members from \citet{vasiliev2021}, based on Gaia DR3 proper motions and parallaxes, a small number of field stars can still be present. Because the available DP1 data around 47\,Tuc cover only the field centred on the cluster itself, we cannot directly measure the impact of any field stars interlopers in the $i$ vs. $(g - i)$ CMD. However, to quantify the remaining field‐star contamination, we defined an ``outer'' control field of the same area as our cluster region, located 2 degrees from the centre of 47\,Tuc, i.e. $(\alpha, \delta) = (12.52, -72.08)$, and selected Gaia DR3 stars there with proper motions, parallaxes, and colours matching 47\,Tuc members.  From this control sample we estimate a $\sim1\%$ contamination rate.  We then followed a similar procedure to \citet{sollima2007} and used the Besan\c{c}on Galaxy Model \citep{czekaj2014}\footnote{https://model.obs-besancon.fr/} to simulate field stars at the same location as our control field, and injected this 1\% of the simulated field stars into the observed CMD. The process was repeated 100,000 times,
recomputing the binary fraction each time subtracting the field single and binary stars from the observed counts. Finally, we adopted the median of the binary fraction distribution  as our ``true'' binary fraction estimate, including the resulting dispersion into our final error budget.

\subsection{Effect of photometric blends}\label{subsec:blends}
Another potential source of uncertainty when measuring $\fbin(q>0.7)$ is photometric blending, which can create artificial binaries, i.e. single stars that, due to blending with neighbour sources, appear brighter and redder like binary stars. To mitigate this effect, we first relied on the metrics provided by the LSST Science Pipeline \citep{developers2025}, which applies an iterative deblending process \citep[see e.g., Sections 4.8 and 4.9 of][for a detailed description]{bosch2018full}. Specifically, we excluded all sources with failed deblending and any with \texttt{g\_blendedness}, \texttt{r\_blendedness}, or \texttt{i\_blendedness} $\geq 0.05$. The blendedness parameter in each band quantifies how much of a source’s flux is affected by neighbours \citep[see e.g., Section 4.9.11 of][for further details]{bosch2018full}. This step ensures that our 47 Tuc sample is already robust against strong blending.

As an additional conservative test, we carried out a simple stellar density-based analysis. For each star we counted the number of neighbours in a narrow annulus between $2$ and $3''$, and compared it with the observed count inside a $2''$ circle. These radii correspond to the worst-case PSF width in our data (from the \texttt{DP1.Visits} table). Stars with inner neighbour counts significantly exceeding the local prediction were flagged as possible blends. Applying the same quality criteria as in Section~\ref{sec:data}, we found that only $\sim 0.5\%$ of stars in the final sample fall into this category and lie within the binary selection region (region B in Figure~\ref{fig:fig2}). We therefore conclude that any residual blending has a negligible effect on our measured binary fraction.

\section{SUMMARY AND DISCUSSION}
\label{sec:conclusion}
We used the coadd photometric catalogue of Rubin Data Preview 1 \citep{rubinDP1dataset2025, rubinDP1paper2025} to analyse the $i$ vs. $g-i$ CMD of the globular cluster 47\,Tuc. The catalogue is obtained coadding together multiple visits for each band (see Section~\ref{sec:data} and \citet{choi2025, wainer2025} for a detailed description) and provide high precision photometry for stars down to the main sequence. Crossmatching Rubin photometric data with Gaia 47\,Tuc cluster members from \citet[][]{vasiliev2021} and selecting well measured stars produced a total of 2206 stars, similar to \cite{choi2025} and \cite{wainer2025}. These stars are located between 18 and 40\,arcmin almost reaching the tidal radius. 

Following the procedure described in Section~\ref{sec:discussion} we find a total of 1308 main sequence stars with masses between $0.60$ and $0.80 M_\odot$, and 25 binary stars with primary masses in the same mass range and mass ratio $q>0.7$. Accounting for residual field stars contamination as discussed in Section~\ref{subsec:field}, we find a binary fraction of $\fbin(q>0.7) = 0.016 \pm 0.005$. As a further check, we also determined the binary fractions analysing the $i$ vs. $r-i$ and $r$ vs. $g-r$ CMDs. We found $\fbin(q>0.7) = 0.018 \pm 0.006$ and $\fbin(q>0.7) = 0.022 \pm 0.007$ for the former and latter, respectively. Hence, different CMDs yield binary fractions consistent within the uncertainties.

Our measured binary fraction in the outskirts of 47\,Tuc differs at 1.9$\sigma$ from the fraction reported by \citet[][$0.005\pm0.003$]{milone2012a} for the region between the core and half-mass radius. Extrapolating our derived binary fractions for $q>0.7$ to all mass-ratios, and assuming a flat mass-ratio distribution, we obtain a total binary fraction of $\fbin(q>0)=0.053 \pm 0.017$, in agreement with the binary fractions in the core determined in \citet[][$\sim 6\%$, their Figure~12]{mullerhoron2025} from Multi Unit Spectroscopic Explorer (MUSE) data. This value is also consistent within the uncertainties with the binary fraction determined in \cite[][between 3 and 8\% depending on the adopted method]{ji2015}, even though a direct comparison is not possible as they provide the binary fractions with $q>0.5$. 

Moreover, while many clusters exhibit a decreasing binary fraction with radius \citep{sollima2007, dalessandro2011, milone2012a, ji2015, milone2016a}, some display flatter trends \citep[e.g. Figure 34 and 35 of][]{milone2012a}. However, those analyses were limited to within two half-light radii. For 47\,Tuc, our Rubin DP1 catalogue extends from $\sim6\,R_h$ out to $\sim12\,R_\mathrm{h}$, allowing us to study binary stars well into the cluster outskirts.

We speculate that a possible interpretation is that in the cluster’s low-density outskirts, where relaxation times are long and dynamical encounters rare, binaries suffer much less disruption and therefore retain a fraction closer to their primordial value. For example, \citet{hurley2007} studied $N$-body simulations of globular clusters and found that the binary population outside the half-mass radius is dominated by primordial binaries, and is roughly constant with radius (see.e.g their Figure~12). 
Additionally, \citet{ivanova2005} found that reproducing today’s core binary fractions requires initial fractions near 100\%. Together, these results suggest dynamical processing reduces binaries in the centre, while the outskirts preserve a fraction closer to the primordial population.

Furthermore, the radial trends in Blue Straggler Star (BSS) populations offer an interesting comparison. \citet{ferraro2004} found that in 47\,Tuc the BSS fraction decrease from the core out to $\sim10$ times the core radius, then rises again toward the cluster outskirts (see their Figure~5). \citet{mapelli2004} then investigated the radial profiles of BSS in 47\,Tuc by means of $N$-body simulations, showing that a significant fraction of BSS found beyond the half‐mass radius could be the products of mass transfer in primordial binaries \cite[see also][for a similar conclusion]{mapelli2006}. We find a qualitatively similar upturn in the overall binary fraction at large radii, further reinforcing the possible connection between BSS formation and the underlying binary population.

Even with this limited commissioning dataset, the successful detection of unresolved binary stars in 47\,Tuc's outskirts underscores the exceptional photometric capability of Rubin for studying stellar clusters. 

\begin{acknowledgement}

This research was funded by the Australian Government through an Australian Research Council Linkage Infrastructure, Equipment and Facilities grant LE220100007 as well as the National Collaborative Research Infrastructure Strategy (NCRIS). This material is based upon work supported in part by the National Science Foundation through Cooperative Agreements AST-1258333 and AST-2241526 and Cooperative Support Agreements AST-1202910 and 2211468 managed by the Association of Universities for Research in Astronomy (AURA), and the Department of Energy under Contract No. DE-AC02-76SF00515 with the SLAC National Accelerator Laboratory managed by Stanford University. Additional Rubin Observatory funding comes from private donations, grants to universities, and in-kind support from LSST-DA Institutional Members. We also thank the anonymous referees for their constructive feedback, which helped improve this work.

\end{acknowledgement}

\bibliography{main}

\end{document}